# Constraining the astrophysical *p* process: Cross section measurement of the $^{84}$Kr($p, \gamma$)$^{85}$Rb reaction in inverse kinematics

A. Palmisano-Kyle,[1,2,3,*] A. Spyrou,[1,2,3] P. A. DeYoung,[4] A. Dombos,[1,2,3,†] P. Gastis,[3,5,‡] O. Olivas-Gomez,[6] C. Harris,[1,2,3] S. Liddick,[2,3,7] S. M. Lyons,[2,3,§] J. Pereira,[1,2,3] A. L. Richard,[2,3,‖] A. Simon,[3,6] M. K. Smith,[2,3] A. Tsantiri,[1,2,3] and R. Zegers[1,2,3]

[1]*Physics Department, Michigan State University, East Lansing, Michigan 48824, USA*
[2]*National Superconducting Cyclotron Laboratory, Michigan State University, East Lansing, Michigan 48824, USA*
[3]*Joint Institute for Nuclear Astrophysics Center for the Evolution of the Elements, University of Notre Dame, Notre Dame, Indiana 46556, USA*
[4]*Physics Department, Hope College, Holland, Michigan 49423, USA*
[5]*Physics Department, Central Michigan University, Mt Pleasant, Michigan 48859, USA*
[6]*Department of Physics and Astronomy, University of Notre Dame, Notre Dame, Indiana 46556, USA*
[7]*Chemistry Department, Michigan State University, East Lansing, Michigan 48824, USA*

One of the biggest questions in nuclear astrophysics is understanding where the elements come from and how they are made. This work focuses on the *p* process, a nucleosynthesis process that consists of a series of photodisintegration reactions responsible for producing stable isotopes on the proton-rich side of stability. These nuclei, known as the *p* nuclei, cannot be made through the well-known neutron-capture processes. Currently *p*-process models rely heavily on theory to provide the relevant reaction rates to predict the final *p*-nuclei abundances and more experimental data is needed. The present work reports on an experiment performed with the SuN detector at the National Superconducting Cyclotron Laboratory, NSCL, at Michigan State University using the ReA facility to measure the $^{84}$Kr($p, \gamma$)$^{85}$Rb reaction cross section in inverse kinematics. The reverse $^{85}$Rb($\gamma, p$)$^{84}$Kr reaction is a branching point in the *p*-process reaction network that was highlighted as an important reaction in sensitivity studies in the production of the $^{78}$Kr *p* nucleus. A new hydrogen gas target was designed and fabricated and a new analysis technique for background subtraction and efficiency calculations of the detector were developed. The experimental cross section is compared to standard statistical model calculations using the NON-SMOKER and TALYS codes.

## I. INTRODUCTION

Heavy element nucleosynthesis is a leading question in the nuclear astrophysics field. The founding nucleosynthesis concepts were first introduced by Burbidge, Burbidge, Fowler, and Hoyle (B2FH) [1] and Cameron [2] in 1957; however, there are still gaps in our knowledge to this day. There are two main neutron-capture processes responsible for creating the stable isotopes of heavy elements known as the *s* (slow) and *r* (rapid) processes [3,4]. However, there are 35 stable proton-rich nuclei ranging in mass from Se ($Z = 34$) to Hg ($Z = 80$) known as the *p* nuclei, which cannot be made through a neutron-capture process [5,6]. *P*-nuclei production is mainly attributed to the *p* process, also known as the $\gamma$ process, which is a series of photodisintegration reactions, such as ($\gamma, n$), ($\gamma, p$), and ($\gamma, \alpha$) reactions [7].

The astrophysical site of the *p* process is still not clear. Some of the proposed sites are Type II supernovae (SNII) [8,9] and Type Ia supernovae [10]. For many decades, the most favored astrophysical site has been SNII when the shock wave passes through the O/Ne layer of the star where the relevant temperatures for the *p* process, from 1.5–3.5 GK, are reached. However, recent studies indicate that the Type Ia scenario may make a more dominant contribution to the *p*-nuclei abundances [11]. The $\nu p$ process in neutrino driven winds of a SNII is thought to also contribute to the production of the *p* nuclei [12].

In addition to astrophysical uncertainties, there are also nuclear physics uncertainties that go into abundance calculations. *P*-process network calculations consist of approximately 20 000 reactions on stable and radioactive isotopes, both of which require nuclear data input. Several measurements have been performed over the last 20 years on stable isotopes [6]. However, to date, there has only been one direct measurement on a radioactive isotope [13]. Sensitivity studies have been performed to identify the most important reactions that affect the final abundances of the *p* nuclei [14–16]. In one such sensitivity study [15], Rauscher *et al.* performed a model-independent investigation to identify important branching points along the *p*-process reaction path. The dominating reactions in this reaction network are ($\gamma$, $n$) reactions; however, for certain isotopes the ($\gamma$, $p$) or ($\gamma$, $\alpha$) reactions begin to dominate [15]. Those isotopes are branching points in the reaction network where experimental measurements are important to identify how the reaction network flows through that region and can subsequently change the final *p*-nuclei abundances. $^{85}$Rb was listed as a branching point where the reaction network could proceed through $^{84}$Rb via the ($\gamma$, $n$) reaction or $^{84}$Kr via the ($\gamma$, $p$) reaction. The reaction pathway though this mass region can affect the final abundance of the $^{78}$Kr *p* nucleus.

---

[*]Present address: Physics Department, University of Tennessee, Knoxville, Tennessee 37921, USA; apalmisa@utk.edu
[†]Present address: Department of Physics and Astronomy, University of Notre Dame, Notre Dame, Indiana 46556, USA.
[‡]Present address: Los Alamos National Laboratory, Los Alamos, New Mexico 87545, USA.
[§]Present address: Pacific Northwest National Laboratory, Richland, Washington 99354, USA.
[‖]Present address: Lawrence Livermore National Laboratory, Livermore, California 94550-9234, USA.

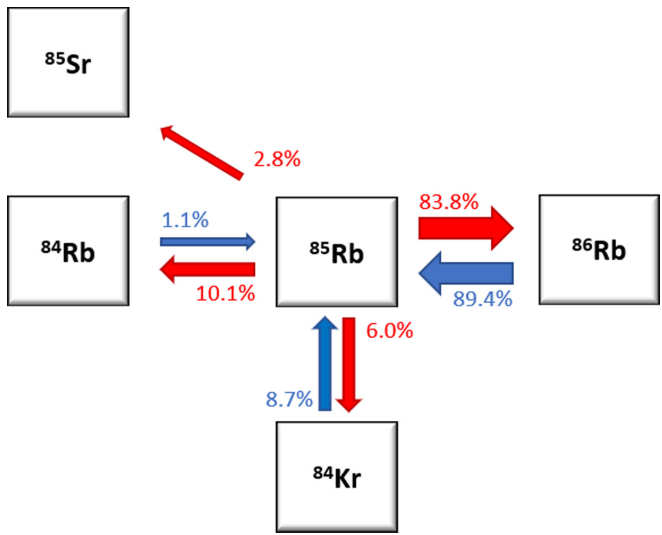

FIG. 1. The theoretical production paths, in blue, and destruction paths, in red, for $^{85}$Rb [16,17]. All reactions have been measured except for the $^{85}$Rb($\gamma$, $p$)$^{84}$Kr branch.

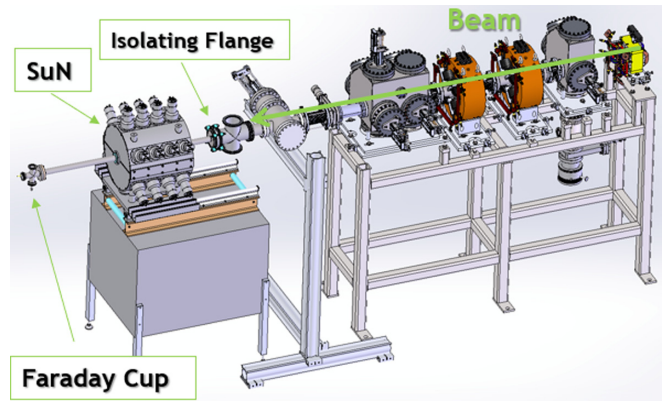

FIG. 2. A diagram of the experimental set up without SuN-SCREEN. Beam comes from the right and enters the setup. The beam current was measured off of the beam pipe which was used as a Faraday cup.

Figure 1 shows the theoretical production and destruction channels of $^{85}$Rb for the *p*-process reaction network [16,17]. The neutron-capture reactions, $^{84}$Rb($n$, $\gamma$)$^{85}$Rb [18], $^{85}$Rb($n$, $\gamma$)$^{86}$Rb [19], and $^{85}$Rb($p$, $n$)$^{85}$Sr [20] have all been previously measured. The final unknown cross section is the $^{85}$Rb($\gamma$, $p$)$^{84}$Kr branch. Therefore, experimentally measuring this reaction will have an important impact on the *p*-process reaction flow in this mass region. The present work reports on a measurement of the $^{84}$Kr($p$, $\gamma$)$^{85}$Rb reaction which can be related to the inverse reaction through the detailed balance theorem [21].

The experimental technique used for the measurement is discussed in Sec. II. The new analysis technique development and its validation with a previously published SuN experiment measuring the $^{90}$Zr($p$, $\gamma$)$^{91}$Nb reaction in Sec. III [22]. Lastly, the results are discussed as well as their impact on the *p*-process reaction flow in the Kr mass region in Sec. IV.

## II. EXPERIMENTAL

The experiment was performed at the National Superconducting Cyclotron Laboratory at Michigan State University using the ReAccelerator (ReA) facility [23]. A beam of $^{84}$Kr was delivered to the experimental setup at four different beam energies: 2.8 MeV/u, 3.1 MeV/u, 3.4 MeV/u, and 3.7 MeV/u. The $^{84}$Kr($p$, $\gamma$)$^{85}$Rb cross section was measured in inverse kinematics with the summing NaI(Tl) (SuN) detector shown in Fig. 2 [24]. SuN is a cylindrical detector of optically isolated NaI crystals 16 inches in length and diameter with a 1.77 inch borehole in the center resulting in nearly $4\pi$ angular coverage. Because SuN has optically isolated segments, the sum-of-segments spectra and the total absorption spectra (TAS) can both be analyzed separately [24]. The TAS sums the $\gamma$ rays on an event-by-event basis and is sensitive to the populated excitation energies. The sum of segments adds the histograms of the individual segments and is sensitive to the individual $\gamma$-ray transitions.

The scintillating cosmic ray eliminating ensemble (SuN-SCREEN) [25] was placed above SuN to help eliminate the cosmic-ray background as an active veto detector. SuN-SCREEN is comprised of nine plastic scintillator bars and to reduce dark noise from the PMTs a coincidence is required between the front and back PMT in the same bar.

A small gas-cell target filled with hydrogen gas was placed in the center of SuN which was a new addition to the experimental setup. It is made of plastic with a tantalum ring on the front face and tantalum foil lining the inside of the cell to shield the beam from the plastic and reduce beam-induced background. The gas cell has a 2-$\mu$m-thick molybdenum foil which was used for the front and back cell window and is approximately 4 cm in length. It was filled with 600 Torr of hydrogen gas for the experiment and the beam loses approximately 0.19–0.22 MeV/u as it goes through the molybdenum foil depending on the initial beam energy.

## III. ANALYSIS AND RESULTS

### A. The $\gamma$-summing technique

The experiment was run in inverse kinematics with a $^{84}$Kr beam impinging on a hydrogen-gas-cell target in the center of SuN. SuN has been used previously in inverse kinematics to successfully measure the $^{27}$Al$(p, \gamma)^{28}$Si reaction cross section [26]. The excited state and the sum peak appear at $E_{c.m.} + Q$, where $E_{c.m.}$ is the center-of-mass energy and $Q$ is the $Q$ value of the reaction. The beam energies ranged from 2.8 MeV/u to 3.7 MeV/u and with a $Q$ value of 7.0 MeV this places the sum peaks in the region of 7–9 MeV. The sum peaks were wider than forward kinematics measurements due to the Doppler shift and energy straggling through the Mo foil and the $H_2$ gas.

### B. Background subtraction

Background subtraction of both beam-induced and room background was very important for this analysis. To isolate cosmic-ray and room background, two subtraction methods were implemented. SuNSCREEN was used as an active cosmic-ray veto detector where events in SuN coming in coincidence with SuNSCREEN were rejected. This coincidence rejection is the first step in the background isolation process.

Secondly, the beam timing structure was used to isolate the remaining room background. The beam was pulsed in 100 $\mu$s pulses separated by 200 ms. The data collection was triggered by the beam pulses creating 100 $\mu$s beam gates. To isolate the remaining room background within the beam gate, a shifted background gate with the same 100 $\mu$s length was also recorded. Beginning 100 ms after the initial beam pulse, this gate recorded only room background events and were subtracted from the beam gates.

After subtracting the cosmic-ray and room backgrounds from the spectra, the beam-induced background from the gas cell needs to be isolated. For this reason, data were taken with the gas cell filled with 600 Torr of hydrogen gas and devoid of hydrogen gas. For the empty-cell runs, the gas cell and the beam pipe were in vacuum. The empty-cell runs were scaled as a function of deposited current and subtracted from the full-cell runs.

After completing the background subtraction, Doppler corrections were applied on a segment-by-segment basis as outlined in Ref [26]. Figure 3 shows the fully Doppler corrected and background subtracted TAS for the highest beam energy for the $^{84}$Kr$(p, \gamma)^{85}$Rb cross section in purple. The black dotted line corresponds to the spectrum taken with the gas cell full of hydrogen gas while the red dotted-dashed line is the scaled empty gas cell spectrum. To arrive at the purple spectra shown in Fig. 3, first an anticoincidence is applied for SuN and SuNSCREEN events. Secondly, a beam gate and background gate are created for all of the data for each beam energy for both the full-cell and empty-cell runs. The empty-cell runs as scaled as a function of beam current listed in Table I and subtracted from the full-cell runs.

The sum peak is still broader than in a forward kinematics experiment due to the Doppler shift and the energy loss of the beam through the gas cell. The energy loss through the cell is the dominant factor in the peak broadening which means the previous analysis techniques for calculating the detector efficiency used in Refs. [22,24] are not applicable here. While the measurement performed in Ref. [22] was also in inverse kinematics, it had a thin solid target which did not affect the detector efficiency. In order to account for the efficiency due to the gas cell, a new method was developed and is outlined in Sec. III C.

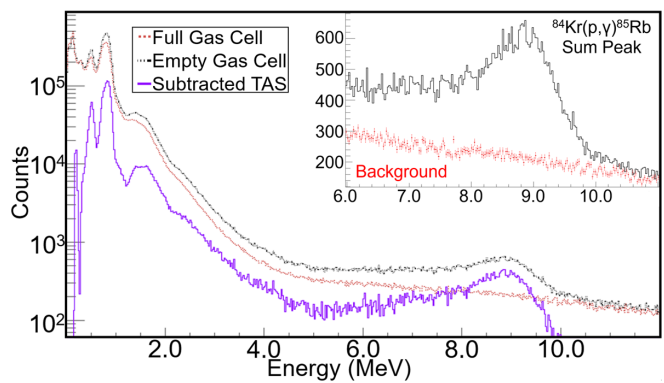

FIG. 3. Total absorption spectra for the $^{84}$Kr$(p, \gamma)^{85}$Rb reaction at the highest beam energy, 3.7 MeV/u. The black dotted line shows the gas cell filled with hydrogen gas, the red dotted-dashed line shows the scaled empty cell spectrum and the purple line is the fully background subtracted and Doppler corrected spectrum. The purple spectrum was used for the remaining analysis and the insert is a closer view of the sum peak above background.

### C. Efficiency validation: $^{90}$Zr$(p, \gamma)^{91}$Nb

SuN's summing efficiency is more complex due to the Doppler broadening of the peaks. When using a thin CH2 target, as was done in [26], the traditional techniques for the efficiency correction are still valid. However, in the present work the use of a gas cell with a relatively thick entrance window caused additional broadening.

A new analysis technique was developed and validated with a previously published data set using SuN in regular kinematics measuring the $^{90}$Zr$(p, \gamma)^{91}$Nb cross section. These data were analyzed with the traditional efficiency technique outlined in Ref. [22] and compared to the new efficiency technique for validation. The new analysis technique consists of a two-step simulation process using both the RAINIER [27] and GEANT4 [28] codes. RAINIER is a statistical model code that simulates the $\gamma$-ray cascades from an excitation energy of a compound nucleus. It outputs a series of $\gamma$-ray cascades which are then input into GEANT4 accounting for the detector response of SuN [27]. The simulated and experimental TAS, sum-of-segments, and multiplicity spectra are compared, and the efficiency is extracted.

TABLE I. The total deposited beam current and number of particles per beam energy.

| Beam energy (MeV) | Current (C) | Number of particles |
|---|---|---|
| 3.7 | $1.36 \times 10^{-06}$ | $1.36 \times 10^{11}$ |
| 3.4 | $1.20 \times 10^{-06}$ | $2.77 \times 10^{11}$ |
| 3.1 | $2.63 \times 10^{-06}$ | $2.64 \times 10^{11}$ |
| 2.8 | $3.92 \times 10^{-06}$ | $9.06 \times 10^{11}$ |

The first step is understanding the $\gamma$-ray cascades from the excited compound nucleus. These cascades define the $\gamma$-ray energy and multiplicity which subsequently impacts the summing efficiency of SuN. RAINIER must be run at the specific excitation energy, $E_x$, that corresponds to the energy of the reaction. The cascades depend on the choice of nuclear level density and $\gamma$-ray strength function within RAINIER.

There are two different level density models in RAINIER that can be chosen: the constant temperature model (CTM) or the back-shifted Fermi gas model (BSFGM) [29,30]. In RAINIER the user can also select different $\gamma$-ray strength functions for various transition types: $E1$, $M1$, and $E2$. $E1$ can be either the standard Lorentzian [31,32], general Lorentzian [33], the Kadmenskij-Markushev-Furman model (KMF) [34], or the Kopecky-Chrien model (KOP). $M1$ transitions are typically described as a standard Lorentzian with or without an up bend at low energies [35]. A systematic investigation of the RAINIER parameters was performed in [36] to find the best match for experimental data for the $^{90}$Zr$(p,\gamma)^{91}$Nb reaction. This reaction was chosen because it was a regular kinematics experiment with a solid thin target and would have a more simplistic simulation process. It was also possible to perform the new efficiency analysis and keep the rest of the original process in Ref. [22] identical. In this way we could isolate the efficiency analysis to validate the new technique. More details on this process are shown in Ref [36].

Once the RAINIER inputs are chosen, the $\gamma$-ray cascades are run through GEANT4 to account for the detector response before comparing the simulations to the experimental data. The number of counts in the sum peak of the simulation compared to the total number of simulated events gives the simulated summing efficiency of the SuN detector for that particular reaction and beam energy. This calculation must be repeated for each beam energy. The cross section was recalculated for the $^{90}$Zr data with the new efficiencies but the remainder of the analysis was unchanged. Figure 4 shows the cross sections for the new technique as blue stars and the original data points are shown as green circles [22]. The two techniques agree within uncertainty and validate the new analysis technique which was then used to calculate the $^{84}$Kr$(p,\gamma)^{85}$Rb cross section.

### D. $^{84}$Kr$(p,\gamma)^{85}$Rb cross section

The newly developed efficiency technique was applied to the $^{84}$Kr$(p,\gamma)^{85}$Rb reaction taking into account modifications due to the use of the gas-cell target. Table II details the energetics of the beam as it passes through various stages of the gas cell. The energy width of the gas cell, $\Delta E$, affects the width of the sum peak and consequently gives a range of available entry state energies for the reaction since the beam can interact at any point within the gas cell. In order to account for this in the efficiency simulations, RAINIER was run with different $E_x$ energies to cover the energy width of the gas cell in increments of 0.1 MeV.

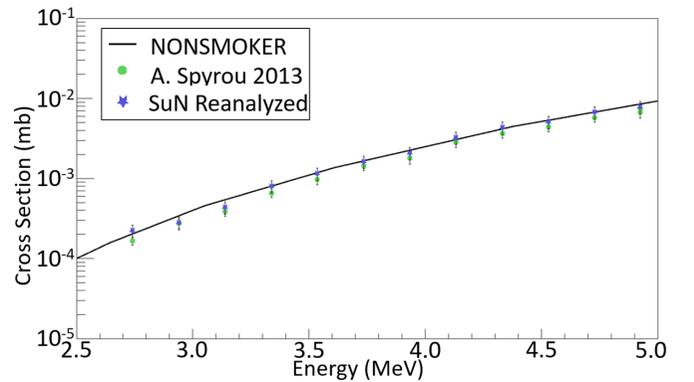

FIG. 4. The previously published SuN data for the $^{90}$Zr$(p,\gamma)^{91}$Nb reaction are displayed as green circles [22], the NON-SMOKER calculations as a black line, and the cross section values calculated with the new efficiency analysis are shown as blue stars.

A $\chi^2$ minimization was run on the simulations to find the overall contribution to the sum peak, or feeding ratio, of each available $E_x$ within the gas cell range. The TAS, sum-of-segments, and multiplicity spectra are all fit simultaneously, and the sum of segments and multiplicity are gated on the sum peak energy range. Example fits for the 3.1 MeV beam energy are shown in Fig. 5 with the simulations shown as dotted red lines and the experimental data shown as black lines.

In order to account for uncertainties from the simulation process, RAINIER was run with 12 different sets of level density and $\gamma$-ray strength function parameters. The analysis was repeated for each parameter set at each beam energy. To extract the simulated detector yield, the feeding ratios were summed and multiplied by the number of simulated events. Once the efficiency was calculated, the final cross section was calculated using the following equation:

$$\sigma = \frac{A}{N_A \xi} \frac{1}{N_b} \frac{I}{\epsilon}, \quad (1)$$

where $A$ is the target mass, $N_A$ is Avogadro's number, $\xi$ is the target thickness, $N_b$ is the number of the beam particles, $I$ is the number of times the reaction is detected, and $\epsilon$ is the detector efficiency. To account for the uncertainty in the energy due to the energy loss of the beam through the gas cell an effective energy was calculated [37]. Within the energy range of the gas cell, the measured cross section could have come from any of the available entry state energies with a higher probability of coming from the higher energy states which have higher cross sections. This effective energy is calculated by

$$E_{\text{eff}} = E_0 - \Delta E + \Delta E \left[ -\frac{\sigma_2}{\sigma_1 - \sigma_2} + \sqrt{\frac{\sigma_1^2 + \sigma_2^2}{2(\sigma_1 - \sigma_2)^2}} \right], \quad (2)$$

where $E_0$, $E_f$, and $\Delta E$ are listed in Table II, $\sigma_1 = \sigma(E_0)$ and $\sigma_2 = \sigma(E_f)$ [38]. The cross sections used to calculate the $E_{\text{eff}}$ were taken from NON-SMOKER [39].

TABLE II. Details of the beam energy as it goes through the gas-cell used for cross-section measurements. The center of mass energy, $E_0$, after the entrance foil, the center of mass energy, $E_f$, after the full gas cell prior to the exit foil, and the energy width, $\Delta E$, of the gas cell.

| Beam energy (MeV) | $E_0$ (MeV) | $E_f$ (MeV) | $\Delta E$ (MeV) |
|---|---|---|---|
| 3.7 | 3.060 | 2.835 | 0.225 |
| 3.4 | 2.758 | 2.535 | 0.223 |
| 3.1 | 2.458 | 2.235 | 0.223 |
| 2.8 | 2.192 | 1.994 | 0.198 |

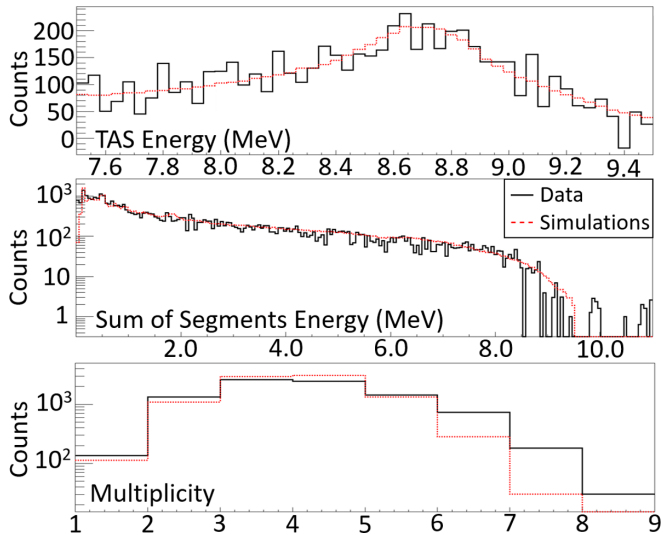

FIG. 5. An example of the $\chi^2$ minimization fits of the TAS, sum of segments and multiplicity spectra for the 3.1 MeV/u beam energy. The red dotted lines are simulations and the black lines are the experimental data.

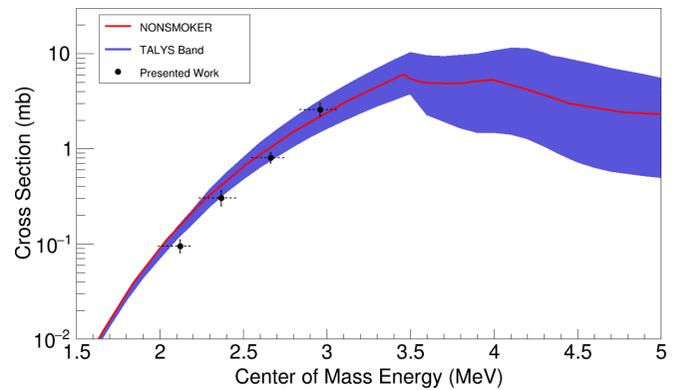

FIG. 6. The measured cross section of the $^{84}$Kr$(p,\gamma)^{85}$Rb reaction are shown as black circles and the associated uncertainty. The dashed line shows the energy width of the gas cell and the data point is presented at the effective energy. The blue band shows the TALYS calculations and the red line shows the NON-SMOKER calculations.

The simulated efficiencies were used to calculate the cross sections with Eq. (1) for each set of inputs and averaged to find an uncertainty in the simulation and fitting processes. The average variation in the final cross section due to the simulations and fitting was 2%. With the technique used here the efficiency of SuN for detecting the full energy peak was 46(4)%. No significant variation was observed within the uncertainty. Uncertainties from the beam current measurement and gas cell pressure were approximately 5% and the statistical uncertainty was 10%.

## IV. DISCUSSION

The final measured cross sections at the effective energy are shown in Fig. 6. The red line corresponds to the NON-SMOKER calculations, the blue band shows the TALYS calculations, and the black circles are the the newly measured cross section values of the $^{84}$Kr$(p,\gamma)^{85}$Rb reaction and its uncertainty at the center of mass effective energy [39,40]. The dashed black line shows the energy range of the gas cell. The blue band is calculated in TALYS by using all of the different level density and $\gamma$-ray strength function parameter combinations for this reaction.

The measured cross sections agree with the theoretical predictions within uncertainty. With the experimental values matching theory it constrains the uncertainties of the nuclear inputs for this reaction in the astrophysical models. Figure 7 shows the *p*-process reaction flow in the Kr mass region. It is typically dominated by the light blue highlighted $(\gamma, n)$

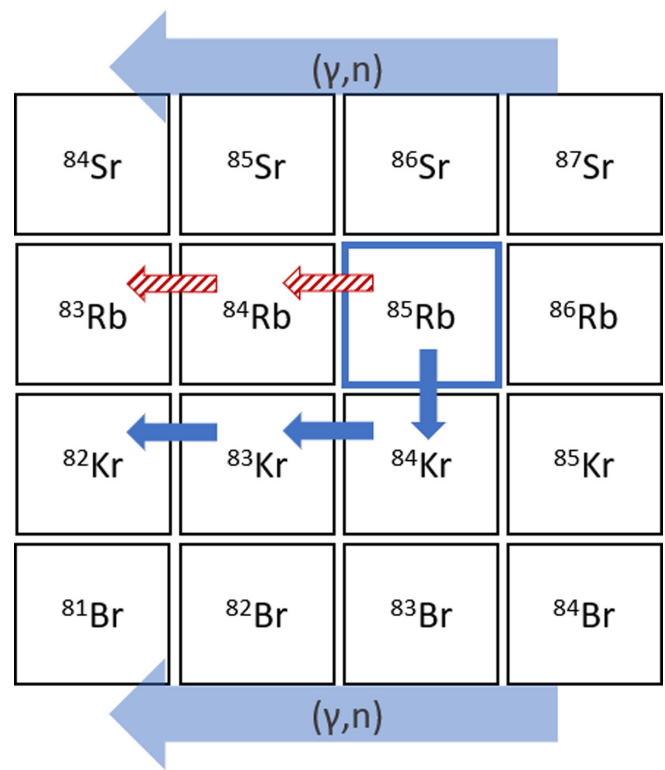

FIG. 7. The mass region around $^{84}$Kr with the predominant $(\gamma, n)$ reactions highlighted in light blue at the top and bottom of the figure. With the $^{84}$Kr$(p,\gamma)^{85}$Rb cross section agreeing with theory, the reaction flow would follow the blue highlighted reaction chain instead of the striped red reaction chain.

TABLE III. The measured cross sections for the $^{84}$Kr$(p, \gamma)^{85}$Rb reaction at the four effective energies and the associated uncertainty.

| $E_{\text{eff}}$ (MeV) | Cross section (mb) | Uncertainty (mb) |
|---|---|---|
| 2.961 | 2.572 | 0.389 |
| 2.663 | 0.801 | 0.099 |
| 2.367 | 0.305 | 0.056 |
| 2.121 | 0.095 | 0.014 |

reactions shown at the top and bottom of the figure until branching points are hit.

Recall that the $^{85}$Rb$(\gamma, p)^{84}$Kr reaction was marked as a branching point between the $(\gamma, n)$ and $(\gamma, p)$ reactions. If the cross section for the $^{84}$Kr$(p, \gamma)^{85}$Rb reaction matches theory, as is shown with this measurement, the reaction network would flow through the chain with the higher predicted reaction rate. Examining the reaction rates predicted by the JINA REACLIB database in the Gamow window temperature range of 2–3.5 GK, the $(\gamma, p)$ chain shown in blue will dominate instead of the $(\gamma, n)$ chain shown in striped red [41]. Therefore the path to the $p$ nucleus, $^{78}$Kr, should mainly go through the krypton chain instead of the rubidium chain.

There has been a recently published value for the cross section of the $^{84}$Kr$(p, \gamma)^{85}$Rb reaction by Lotay *et al.* [13]. They have quoted the cross section as $94^{+64}_{-30}$ $\mu$b at an effective center of mass energy of 2.435 MeV which does not agree with the cross sections presented in this work in Table III. An independent measurement on this cross section is needed to resolve the discrepancy. Despite the disagreement between these measurements, it remains true that the $(\gamma, p)$ chain will dominate.

## V. CONCLUSIONS

The $^{84}$Kr$(p, \gamma)^{85}$Rb cross section was directly measured using the SuN detector in inverse kinematics. A new hydrogen gas-cell target was designed and tested for this experiment as well as a new efficiency analysis technique. The new efficiency analysis technique was validated with the previously published $^{90}$Zr$(p, \gamma)^{91}$Nb cross section measured with SuN [22]. This measurement confirms the direction of the $p$-process reaction flow which feeds the creation of the $^{78}$Kr $p$ nucleus as proceeding via the krypton chain instead of the rubidium chain. This experimental technique can be used for future measurements to help further constrain the p-process reaction network by measuring currently unknown $p$-process cross sections on radioactive isotopes.

## ACKNOWLEDGMENTS

The authors would like to acknowledge the support of the ReA3 accelerator team with optimizing the beam delivery and setup. The work was supported by the National Science Foundation under Grants No. PHY 1613188 (Hope College), PHY 1102511 (NSCL), PHY 1913554 (Windows on the Universe: Nuclear Astrophysics at the NSCL), PHY 1430152 (Joint Institute for Nuclear Astrophysics). This material is based upon work supported by the U.S. Department of Energy, National Nuclear Security Administration through Grant No. DOE-DE-NA0003906, Award No. DE-NA0003180 (Nuclear Science and Security Consortium) and Grant No. DE-FG02-96ER40963 from the Department of Energy, Office of Science, Office of Nuclear Physics.

[1] E. Burbidge, G. Burbidge, W. Fowler, and F. Hoyle, Rev. Mod. Phys. **29**, 547 (1957).
[2] A. G. W. Cameron, Astron. Soc. Pac. **69**, 201 (1957).
[3] F. Käppeler, R. Gallino, S. Bisterzo, and W. Aoki, Rev. Mod. Phys. **83**, 157 (2011).
[4] M. Arnould, S. Goriely, and K. Takahashi, Phys. Rep. **450**, 97 (2007).
[5] M. Arnould and S. Goriely, Phys. Rep. **384**, 1 (2003).
[6] T. Rauscher, N. Dauphas, I. Dillmann, C. Fröhlich, Z. Fülöp, and Gy. Gyürky, Rep. Prog. Phys. **76**, 066201 (2013).
[7] B. Meyer, Annu. Rev. Astron. Astrophys. **32**, 153 (1994).
[8] M. Rayet, M. Arnould, and N. Prantzos, Astron. Astrophys. **227**, 271 (1990).
[9] T. Rauscher, A. Heger, R. Hoffman, and S. Woosley, Astrophys. J. **576**, 323 (2002).
[10] C. Travaglio, F.K. Röpke, R. Gallino, and W. Hillebrandt, Astrophys. J. **739**, 93 (2011).
[11] C. Travaglio, T. Rauscher, A. Heger, M. Pignatari, and C. West, Astrophys. J. **854**, 18 (2018).
[12] M. Kostur, M. Schindler, P. Talkner, and P. Hanggi, Phys. Rev. Lett. **96**, 014502 (2006).
[13] G. Lotay, S. A. Gillespie, M. Williams, T. Rauscher, M. Alcorta, A. M. Amthor, C. A. Andreoiu, D. Baal, G. C. Ball, S. S. Bhattacharjee, H. Behnamian, V. Bildstein, C. Burbadge, W. N. Catford, D. T. Doherty, N. E. Esker, F. H. Garcia, A. B. Garnsworthy, G. Hackman, S. Hallam, K. A. Hudson, S. Jazrawi, E. Kasanda, A. R. L. Kennington, Y. H. Kim, A. Lennarz, R. S. Lubna, C. R. Natzke, N. Nishimura, B. Olaizola, C. Paxman, A. Psaltis, C. E. Svensson, J. Williams, B. Wallis, D. Yates, D. Walter, and B. Davids, Phys. Rev. Lett. **127**, 112701 (2021).
[14] T. Rauscher, N. Nishimura, R. Hirschi, and G. Cescutti, Mon. Not. R. Astron. Soc. **463**, 4153 (2016).
[15] T. Rauscher, Phys. Rev. C **73**, 015804 (2006).
[16] W. Rapp, J. Gorres, M. Wiescher, H. Sschatz, and F. Kaeppeler, Astrophys. J. **653**, 474 (2006).
[17] https://exp astro.de/fluxes/, P-process flux calculator.
[18] C. Plaisir, F. Gobet, M. Tarisien *et al.*, Eur. Phys. J. A **48**, 68 (2012).
[19] N. Dudey, R. Heinrich, and A. Madson, J. Nucl. Energy **24**, 181 (1970).
[20] G. G. Kiss, Gy. Gyürky, A. Simon, Zs. Fülöp, E. Somorjai, and T. Rauscher, in *Capture Gamma-Ray Spectroscopy and Related*

*Topics: Proceedings of the 13th International Symposium on Capture Gamma-Ray Spectroscopy and Related Topics*, edited by J. Jolie, A. Zilges, N. Warr, and A. Blazhev, AIP Conf. Proc. No.1090 (AIP, New York, 2009), p. 476.

[21] J. Blatt and V. Weisskopf, *Theoretical Nuclear Physics* (Wiley, New York, 1952).

[22] A. Spyrou, S. J. Quinn, A. Simon, T. Rauscher, A. Battaglia, A. Best, B. Bucher, M. Couder, P. A. DeYoung, A. C. Dombos, X. Fang, J. Gorres, A. Kontos, Q. Li, L. Y. Lin, A. Long, S. Lyons, B. S. Meyer, A. Roberts, D. Robertson, K. Smith, M. K. Smith, E. Stech, B. Stefanek, W. P. Tan, X. D. Tang, and M. Wiescher, Phys. Rev. C **88**, 045802 (2013).

[23] A. Lapierre, S. Schwarz, K. Kittimanapun, J. Rodriguez, C. Sumithrarachchi, B. Barquest *et al.*, Nucl. Instrum. Methods Phys. Res. B **317**, 399 (2013).

[24] A. Simon, S.J. Quinn, A. Spyrou, A. Battaglia, Nucl. Instrum. Methods Phys. Res. A **703**, 16 (2013).

[25] E. Klopfer, J. Brett, P. A. DeYoung, A. Dombos, S. Quinn, A. Simon, and A. Spyrou, Nucl. Instrum. Methods Phys. Res. A **788**, 5 (2015).

[26] S. J. Quinn, A. Spyrou, A. Simon, and A. Battaglia, Nucl. Instrum. Methods Phys. Res. A **757**, 62 (2014).

[27] L. Kirsch and L. Berstein, Nucl. Instrum. Methods Phys. Res. A **892**, 30 (2018).

[28] S. Agostinelli, J. Allison, K. Amako, J. Apostolakis, H. Araujo, P. Arce *et al.*, Nucl. Instrum. Methods Phys. Res. A **506**, 250 (2003).

[29] A. Gilbert and A. Cameron, Can. J. Phys. **43**, 1446 (1965).

[30] W. Dilg, W. Schantl, H. Vonach, and M. Uhl, Nucl. Phys. A **217**, 269 (1973).

[31] D. Brink, Nucl. Phys. **4**, 215 (1957).

[32] P. Axel, Phys. Rev. **126**, 671 (1962).

[33] J. Kopecky and M. Uhl, Phys. Rev. C **41**, 1941 (1990).

[34] S. G. Kadmenskij, V. P. Markushev, and V. I. Furman, Yad. Fiz. (USSR) **37**, 277 (1983).

[35] A. C. Larsen, J. E. Midtbo, M. Guttormsen, T. Renstrom, S. N. Liddick, A. Spyrou, S. Karampagia, B. A. Brown, O. Achakovskiy, S. Kamerdzhiev, D. L. Bleuel, A. Couture, L. C. Campo, B. P. Crider, A. C. Dombos, R. Lewis, S. Mosby, F. Naqvi, G. Perdikakis, C. J. Prokop, S. J. Quinn, and S. Siem, Phys. Rev. C **97**, 054329 (2018).

[36] A. Palmisano, Constraining the P Process: Cross Section Measurement of $^{84}$Kr$(p,\gamma)^{85}$Rb, Ph.D. thesis, Michigan State University (2021).

[37] C. Illiadis, *Nuclear Physics of Stars* (Wiley-VCH Verlag GmbH & Co. KGaA, Weinheim, Germany, 2007).

[38] C. E. Rolfs and W. S. Rodney, *Cauldrons in the Cosmos: Nuclear Astrophysics* (The University of Chicago Press, Chicago, 1988).

[39] T. Rauscher and F.-K. Thielemann, At. Data Nucl. Data Tables **75**, 1 (2000).

[40] A. Koning, S. Hilaire, and S. Goriely, TALYS-1.8: A nuclear reaction program, 1st ed. (2015).

[41] R. H. Cyburt, A. M. Amthor, R. Ferguson, Z. Meisel, K. Smith, S. Warren, A. Heger, R. D. Hoffman, T. Rauscher, A. Sakharuk, H. Schatz, F. K. Thielemann, and M. Wiescher, Astrophys. J. Suppl. Series **189**, 240 (2010).